\documentclass[12pt,a4paper]{amsart}

\usepackage{amsmath,amsfonts,amssymb}

\newcommand{\Z}{\mathbb Z}
\newcommand{\C}{\mathbb C}

\newcommand{\T}{\mathcal{T}}
\def\d{\partial}
\def\CP1{\mathbb{C}\mathrm{P}^1}
\def\un{{1\!\! 1}}
\def\r{\mathfrak{r}}
\def\s{\mathfrak{s}}

\newtheorem{theorem}{Theorem}
\newtheorem{proposition}[theorem]{Proposition}
\newtheorem{lemma}[theorem]{Lemma}

\theoremstyle{definition}

\newtheorem{definition}[theorem]{Definition}
\newtheorem{example}[theorem]{Example}
\newtheorem{remark}[theorem]{Remark}

\title[Dubrovin--Zhang hierarchies]{A polynomial bracket for the Dubrovin--Zhang hierarchies}

\author{A. Buryak}

\address{A.~Buryak:\newline
Department of Mathematics,
University of Amsterdam, \newline
P.~O.~Box 94248, 1090 GE Amsterdam, 
The Netherlands\newline 
\indent and\newline
Department of Mathematics, Moscow State University,\newline
Leninskie gory, 119992 GSP-2 Moscow, Russia}
\email{a.y.buryak@uva.nl, buryaksh@mail.ru}

\author{H. Posthuma}

\address{H.~Posthuma:\newline
Department of Mathematics,
University of Amsterdam, \newline
P.~O.~Box 94248, 1090 GE Amsterdam, 
The Netherlands}

\email{h.b.posthuma@uva.nl}

\author{S. Shadrin}

\address{S.~Shadrin:\newline
Department of Mathematics,
University of Amsterdam, \newline
P.~O.~Box 94248, 1090 GE Amsterdam, 
The Netherlands}
\email{s.shadrin@uva.nl}

\begin{document}

\begin{abstract}
We define a hierarchy of Hamiltonian PDEs associated to an arbitrary tau-function in the semi-simple orbit of the Givental group action on genus expansions of Frobenius manifolds. We prove that the equations, the Hamiltonians, and the bracket are weighted-homogeneous polynomials in the derivatives of the dependent variables with respect to the space variable. 

In the particular case of a conformal (homogeneous) Frobenius structure, our hierarchy coincides with the Dubrovin-Zhang hierarchy that is canonically associated to the underlying Frobenius structure. Therefore, our approach allows to prove the polynomiality of the equations, Hamiltonians and one of the Poisson brackets of these hierarchies, as conjectured by Dubrovin and Zhang. 
\end{abstract}

\maketitle

\tableofcontents

\section{Introduction}

In this paper, we translate some basic notions of the fundamental work of Dubrovin and Zhang~\cite{DubZha2} on bi-Hamiltonian integrable hierarchies associated to Frobenius manifolds~\cite{DubZha2} into the language 
of cohomological field theories, Givental theory, and the related topology of the moduli space of curves. 
From this point of view, Givental's group action on cohomological field theories gives us a new tool to study Dubrovin-Zhang hierarchies, and we use it to establish one of the key properties of the hierarchies conjectured by Dubrovin and Zhang: the polynomiality of one of the Hamiltonian structures.

We refer to the papers of Dubrovin and Zhang~\cite{DubZha1,DubZha2} (see also the expositions of some parts of their theory in~\cite{Get} and~\cite{Ros}) and to a number of papers on Givental's theory~\cite{FabShaZvo,Giv1,Giv2,Lee1,Lee2,LeeVak,Sha} for the necessary general background that we will be able to recall only briefly in this paper. 

\subsection{Dubrovin-Zhang construction and polynomiality} Let us explain the main problem that we address in this paper. Dubrovin and Zhang~\cite{DubZha1,DubZha2} were working on a classifications project for a special class of 1+1 hierarchies that would conjecturally include many interesting hierarchies of this type. Their approach is based on a number of conjectures (in some cases, proved) identifying Gromov-Witten potentials of some target varieties as tau-functions of some hierarchies of KdV-type. 

The construction of Dubrovin and Zhang consists of several steps. First, there is a canonical relation between dispersionless bi-Hamiltonian tau-symmetric hierarchies of hydrodynamic type and semi-simple conformal Frobenius manifolds (that is, semi-simple Frobenius manifolds equipped with an Euler vector field). Second, imposing the Virasoro constrains as an axiom, Dubrovin and Zhang find a unique quasi-Miura transformation that turns the dispersionless hierarchy into a dispersive one. The tau-cover of the resulting dispersive hierarchy has a distinguished solution called \emph{topological} that is conjectured to be the Gromov-Witten potential\footnote{Gromov-Witten theory serves us as just one of the motivating examples, where the objects that we consider do arise in a natural way. Therefore we systematically ignore throughout the paper the subtlety related to the fact that Gromov-Witten potential take values in the Novikov ring rather than in $\C$.} of some target variety $X$ in the case we have started with the Frobenius manifold structure determined by the quantum cohomology of $X$.

The term ``quasi-Miura transformation'' refers to a Miura-type transformation that is not necessarily a polynomial in the derivatives of the dependent variables, but rather a rational function. Exactly this non-polynomiality is the source of problems in the Dubrovin-Zhang construction. The dispersionless hierarchy is polynomial in the derivatives, namely, all its Hamiltonians, equations, and both Poisson brackets are polynomials. On the other hand, all ingredients of the resulting dispersive hierarchy appear to be merely rational functions. In some sense, the canonical nature of the Dubrovin-Zhang construction, in particular, the fact that the quasi-Miura transformation is determined unambiguously by the axiom of Virasoro constraints, allows to control completely the resulting hierarchy. In particular, Dubrovin and Zhang conjectured that the Hamiltonians, the equations, and the brackets are polynomials in the derivatives. In fact, the paper \cite{DubZha2} contains a  proof of the polynomiality of the Hamiltonians and the equations, but  Boris Dubrovin has recently informed us that, unfortunately, there is a gap in their argument.


\subsection{Givental theory} There is another ca\-no\-nical genus expansion of a semi-simple conformal Frobenius manifold. It was given by Givental in terms of the quantization of a group action on the space of Frobenius manifolds~\cite{Giv1,Giv2,Giv3}. Dubrovin and Zhang proved in~\cite{DubZha2} that the topological tau-function that they constructed coincides with the Givental formula. On the other hand, a result of Teleman \cite{Tel} on the classification of semi-simple weighted homogeneous cohomological field theories implies the following: if the quantum cohomology of a target variety determines an analytic semi-simple Frobenius structure, then the full descendant Gromov-Witten potential must coincide with the Givental formula. Therefore, in this setting the conjecture of Dubrovin and Zhang that the topological tau-function of their hierarchy coincides with the corresponding full descendant Gromov-Witten potential is true.

We restrict our attention to a full descendant Gromov-Witten potential, or, more generally, any formal power series in the semi-simple orbit of the quantized Givental group action. If we forget about homogeneity and therefore the Euler vector field, we lose the bi-Hamiltonian structure associated to the underlying Frobenius manifold. However, we still can define some pieces of the structure of the hierarchy purely in terms of this formal power series. This includes the Hamiltonians, equations, and one Poisson bracket of the dispersionless hierarchy, together with a weakened version of a quasi-Miura transformation. With this transformation we can therefore define the Hamiltonians, equations, and one  bracket of the full dispersive hierarchy. A weak quasi-Miura transformation simply means that in the non-homogeneous case we have no control on non-polynomial nature of the transformation that we construct, and we only know, by the result of Dubrovin--Zhang, that it turns into a rational function in the points where an Euler vector field can be introduced.

\subsection{Group action on ingredients of the hierarchy}
We see that by dropping the homogeneity condition, we loose a part of the structure. However, we gain a new tool --- the quantized action of the Givental group. It acts on some special kind of formal power series and its action can be translated into the action on those ingredients of the hierarchy that can be reconstructed from topological tau-functions without a usage of the Euler vector field. 

Of course, we cannot say anything about weak quasi-Miura transformations, since it is even not clear in which class of functions we have to look for its deformations, but we still can compute the infinitesimal action of the Givental group on Hamiltonians, equations, and a bracket of the full dispersive hierarchy associated to a particular point in the semi-simple orbit of the Givental group. It is an amazing computation, quite difficult in many places, and it has a remarkable outcome: the deformation formulas imply that if Hamiltonians, equations, and a bracket that we deform are polynomials at one point in the orbit, they remain to be polynomials in the whole orbit. 

There is indeed one point in the orbit of the Givental group, where everything can be computed explicitly and the polynomiality of all key structures is clear. It is the Gromov-Witten potential of $n$ points, or, in other words, the product of $n$ copies of the topological tau-function of the KdV hierarchy. 

In this way we generalize the conjecture of Dubrovin and Zhang on polynomiality of the Hamiltonians, the equations, and one of the brackets to the case of non-homogeneous Frobenius structures, and we prove it in the more general settings of non-homogeneous Frobenius structures. However, we have to mention that the second bracket is so far completely out of reach for our methods since its definition heavily uses the Euler vector field, which is not well compatible with the Givental group action.

\subsection{Organization of the paper} 

In Section~\ref{sec:action} we recall the key formulas for the Givental group action on the space of tame partition functions associated to Frobenius manifolds. In Section~\ref{sec:hierarchy} we explain how to write down the principal and the full hierarchy associated to an arbitrary tame partition function. In the homogeneous case, it is simply an explanation how to reproduce different ingredients of the Dubrovin-Zhang construction starting from a topological tau-function. 

The equations and the Hamiltonians of the full Dubrovin-Zhang hierarchy are expressed in terms of functions $\Omega_{\alpha,p;\beta,q}$ that are, roughly speaking, the second derivatives of the logarithm of the partition function we have started with.
In Section~\ref{sec:def-omega} we compute the formulas for the infinitesimal deformation of $\Omega_{\alpha,p;\beta,q}$ with respect to the Lie algebra of the Givental group.

The main property of the Poisson bracket is that it turns Hamiltonians into the equations. This allows us to compute an infinitesimal deformation formula for the bracket in Section~\ref{sec:def-bracket}. It is the most complicated computation in the paper. In Section~\ref{sec:unique} we state a uniqueness result that implies that we indeed have deformed the Dubrovin-Zhang canonical bracket (rather than that we have found a new one).

Though the formulas for the infinitesimal deformations that we obtains are fairly complicated, it is enough to look into their structure in order to conclude that they preserve the homogeneous polynomiality of the deformed objects. We discuss that in Section~\ref{sec:conclusions}, and, together with deformations formulas themselves, it is the main result of our paper.

\subsection{Acknowledgements}

We thank G.~Carlet, B.~Dubrovin, S.~Igonin, J.~van de Leur, and D.~Zvonkine for helpful remarks and discussions.


\section{The action of the Givental group} \label{sec:action}
In this section we briefly recall the definition of the Givental group and its action on the so-called tame partition functions.

\subsection{Tame partition functions} 
Let $V$ be a vector space of dimension $s$ equip\-ped with a scalar product $\left<~,~\right>$. We fix an orthonormal basis $e_\alpha$, 
$\alpha=1,\dots, s$ and write $\un$ for the element $\sum_{\alpha=1}^s e_\alpha$ in $V$. Next we consider the vector space $V\otimes \C[z]$ and write 
$t=\sum_{\alpha,k}t_{\alpha,k}e_\alpha z^k$ for a generic element of it. 

We shall consider partition functions in the variables $\hbar$ and $t_{\alpha,k}$, $\alpha=1,\dots,s$, $k=0,1,\dots$, of the form
\begin{equation}
\label{potential}
Z(t_{0},t_{1},\dots)=\exp\left(\sum_{g=0}^\infty \hbar^{g-1}F_g(t_{0},t_{1},\dots)\right).
\end{equation}
Here we assume that $\hbar\log Z$ is an analytic function in the variables $t_0=\{t_{1,0},\dots,t_{s,0}\}$ and a formal power series in $\hbar$ and $t_{\alpha,k}$, $\alpha=1,\dots,s$, $k\geq 1$. An example of such a partition function is the generating function of Gromov--Witten invariants of a target variety, or, more generally, the partition function of a cohomological field theory (modulo some convergence issues that are still important in these cases, since we need to check the analyticity).

For such a partition function, we define
\begin{equation}
\label{ham}
\Omega^{[0]}_{\alpha,p\beta,q}:=\frac{\partial^2 F_0}{\partial t_{\alpha,p}\partial t_{\beta,q}},
\end{equation}
and introduce recursively the formal vector fields
\begin{align}
\label{vectorfield}
O_{\alpha,0}&:=\frac{\partial}{\partial t_{\alpha,0}},\\ \notag
O_{\alpha,k}&:=\frac{\partial}{\partial t_{\alpha,k}}-\sum_{i=0}^{k-1}\sum_\beta\Omega^{[0]}_{\alpha,i,\beta,0}O_{\beta,k-i-1}, && k\geq 1.
\end{align}
An important regularity condition on the partition functions is given by \emph{tameness}. In Gromov--Witten theory this property expresses the fact that by the factorization property the potential satisfies an infinite number of equations by pulling back high enough powers of $\psi$-classes on moduli spaces of stable curves, which vanish for dimensional reasons, cf.~\cite{EguXio,Get,Giv2}. We express this property as follows:
\begin{definition}
\label{tame}
A partition function $Z$ is said to be \emph{tame} if 
\begin{align}
\label{trr}
O_{\alpha,k}\left(\frac{\partial^2F_0}{\partial t_{\alpha,p}\partial t_{\beta,q}}\right)&=0,&& k>0, \alpha=1,\dots,s;\\ \label{eq:trr-higher-gen}
O_{\alpha,k}\left(F_g\right)&=0, &&  g\geq 1, k> 3g-2, \alpha=1,\dots,s.
\end{align}
\end{definition}

For example, the topological recursion relation for the $g=0$ potential (TRR-$0$) is equivalent to the equation~\eqref{trr} for $k>0$ in the definition above. In the framework of Frobenius manifolds, it implies the associativity of the multiplication on $V$ and can be used to introduce descendants starting from a prepotential on the small phase space.

Besides these relations, we shall assume that $F_g$, $g\geq 0$, satisfies the \emph{string equation}:
\begin{equation}
\label{string-equation}
\frac{\partial F_g}{\partial t_{\un,0}}=\sum_{\nu,k}t_{\nu,k+1}\frac{\partial F_g}{\partial t_{\nu,k}} +\frac{\delta_{g,0}}{2}\sum_{\alpha}t_{\alpha,0}^2.
\end{equation}
In the case of $g=0$, this equation is related to the existence of a unit vector field on the underlying Frobenius manifold.


\subsubsection{Genus $0$}

In genus $0$, some geometrical meaning of the tameness condition is given by the following proposition.
\begin{proposition}
For a tame potential $F_0$, the vector fields $O_{\alpha,k}$, $k\geq 1$, $\alpha=1,\dots,s$, are in involution: $[O_{\alpha,k},O_{\beta,l}]=0$.
\end{proposition}

\begin{proof}
Indeed, the coefficients of the vector fields $O_{\alpha,k}, O_{\beta,l}$ are polynomials in $\Omega_{\alpha,p;\beta,q}$. Equation~\eqref{trr} implies that the derivatives of coefficient of $O_{\alpha,k}$ with respect to $O_{\beta,l}$ (and vice versa) are equal to zero. Therefore, the commutator is also equal to zero.
\end{proof}

It follows that the vector fields $O_{\alpha,k}$ for $k\geq 1$ define a foliation of codimension $s=\dim(V)$. By condition~\eqref{trr} the functions $\Omega^{[0]}_{\alpha,p,\beta,q}$ are constant along the leaves and can be written as functions of $s$ variables in coordinates adapted to the foliation. This can be done explicitly by the coordinate transformation $t_{\mu,0}\mapsto v_{\alpha}(t)$, where 
\begin{equation}
\label{coord}
v_\alpha(t):=\Omega^{[0]}_{\alpha,0;\un,0}(t_0,t_1,\dots)=\frac{\partial^2 F_0}{\partial t_{\alpha,0}\partial t_{\un,0}}(t_0,t_1,\dots).
\end{equation}
Differentiating the string equation \eqref{string-equation}, one finds that
\begin{equation}\label{eq:v-alpha-0}
v_\alpha=t_{\alpha,0}+\sum_{\nu,k}t_{\nu,k+1}\Omega^{[0]}_{\alpha,0;\nu,k}.
\end{equation}

\begin{proposition} 
We have:
\begin{equation}\label{eq:dep-on-v}
\Omega^{[0]}_{\alpha,p,\beta,q}(t_0,t_1,\ldots)=\Omega^{[0]}_{\alpha,p;\beta,q}(v,0,0,\dots).
\end{equation}
\end{proposition}

\begin{proof}
Indeed, by the previous equation both sides agree when $t_{\alpha,k}=0$, $k\geq 1$, and are constant in the direction of the vector fields $O_{\alpha,k}$, $k\geq 1$.
\end{proof}


\subsubsection{Higher genera}

An analogue of equation~\eqref{eq:dep-on-v} exists for any $g\geq 1$ and is called the $3g-2$ property~\cite{EguXio,DubZha2,Get,Giv2}. 

Let us fix $g\geq 1$. The vector fields $O_{\alpha,k}$ for $k\geq 3g-1$ define a foliation of codimension $s(3g-2)$. By condition~\eqref{eq:trr-higher-gen} the function $F_g$ is constant along the leaves and can be written as functions of $s(3g-2)$ variables in coordinates adapted to the foliation. This can be done explicitly by the coordinate transformation $t_{\mu,k}\mapsto v_{\alpha,k}(t)$, $k\leq 3g-2$, where 
\begin{equation}
v_{\alpha,k}(t):=\frac{\d^k v_\alpha}{\d t_{\un,0}^k}
=\frac{\d^k }{\d t_{\un,0}^k}\Omega^{[0]}_{\alpha,0;\un,0}(t_0,t_1,\dots).
\end{equation}

Differentiating further equation~\eqref{eq:v-alpha-0} and using the string equation~\eqref{string-equation}, one finds that
\begin{equation} \label{eq:formula-v-alpha-k}
v_{\alpha,k}=\delta_{k,1}+t_{\alpha,k}+\sum_{\nu,m}t_{\nu,k+m+1}\Omega^{[0]}_{\alpha,0;\nu,m}+O(t^2).
\end{equation}
In these coordinates we have the following description of $F_g$.
\begin{proposition}
There exist functions $P^{[g]}_0,\dots P^{[g]}_{3g-2}$ of $3g-1$ variables such that
\begin{align} \label{eq:3g-2}
 F_g&(t_0,t_1,\dots) \\ \notag
& =F_g \left( P^{[g]}_0(v_0,\dots,v_{3g-2}),\dots,P^{[g]}_{3g-2}(v_0,\dots,v_{3g-2}),0,0,\dots\right).
\end{align}
\end{proposition}

\begin{proof} First of all, we observe that tameness in genus $0$ implies that $O_{\alpha,k}v_{\beta,m}=0$ for $k\geq m+1$. Therefore, both the left hand side and the right hand side of equation~\eqref{eq:3g-2} are constant along $O_{\alpha,k}$, $k\geq 3g-1$. 

Let us choose functions $P^{[g]}_i$, $i=0,\dots,3g-2$ to be the inverse map to $\{t_{\alpha,k}\}_{k\leq 3g-2} \mapsto \{v_{\alpha,k}\}_{k\leq 3g-2}$ given by equation~\eqref{eq:formula-v-alpha-k} restricted to the subspace $t_{\alpha,k}=0$, $k\geq 3g-1$. Since the foliation spanned by $O_{\alpha,k}$, $k\geq 3g-1$, is transversal to this subspace, we obtain the equation~\eqref{eq:3g-2} on the whole space of variables.
\end{proof}


\subsection{The Givental group and Lie algebra} \label{sec:Givental}

In \cite{Giv2}, Givental introduced the action of the twisted loop group $L^{(2)}GL(V)$ on the space of tame partition functions.\footnote{In fact, here we abuse a little bit the terminology. The twisted loop group action includes a translation of variables $t_{0,\alpha}$, see~\cite{FabShaZvo} for a detailed discussion. So, in general, to a particular partition function one can only apply a group element in a small enough neighborhood of the unit.}. Here we shall use, following~\cite{Lee1,FabShaZvo}, the infinitesimal action of its Lie algebra on the space of tame partition functions
satisfying the string equation~\eqref{string-equation}. For this we use the Birkhoff factorization of the loop group and introduce the Lie algebras
\begin{equation}
\mathfrak{g}_\pm:=\left\{ u(z):=\sum_{k>0}u_kz^{\pm k},~u_k\in{\rm End}(V),~u(-z)^t+u(z)=0\right\}.
\end{equation}
In general we shall write $\r$ for a generic element of $\mathfrak{g}_+$, which is traditionally called \emph{the upper triangular subalgebra}, and $\s$ for an element in $\mathfrak{g}_-$, \emph{the lower triangular subalgebra}. 

Concretely, the upper triangular subalgebra is given by formal power series $\r=\sum_{\ell\geq 1} \r_\ell z^\ell\in\operatorname{End}(V)[[z]]$, where $\r_\ell$ is selfadjoint for $k$ odd and skew-selfadjoint for $k$ even.
Such an element acts on a partition function by the second order differential operator computed in~\cite{Lee1}:
\begin{align} \label{eq:r-action}
\hat{\r} :=&
-\sum_{\begin{smallmatrix}\ell\geq 1 \\ \mu\end{smallmatrix}}
(\r_\ell)^{\mu}_{\un} \frac{\partial}{\partial t_{\mu,\ell+1}}
+\sum_{\begin{smallmatrix}d\geq 0,\ell\geq 1 \\ \mu,\nu\end{smallmatrix}}
(\r_\ell)^{\mu}_{\nu} t_{\nu,d}\frac{\partial}{\partial t_{\mu,d+\ell}} \\ \notag 
& + \frac{\hbar}{2}
\sum_{\begin{smallmatrix}i,j\geq 0 \\ \mu,\nu\end{smallmatrix}}
(-1)^{i+1}(\r_{i+j+1})^{\mu,\nu}\frac{\partial^2}{\partial t_{\mu,i}\d t_{\nu,j}}.
\end{align}
For $\s=\sum_{\ell\geq 1}\s_\ell z^{-\ell}$, an element of the lower triangular subalgebra, we have the first order differential operator computed in~\cite{Lee1}:
\begin{align} \label{eq:s-action}
\hat{\s} :=& 
-\frac{1}{2\hbar} (\s_3)_{\un,\un}
+\frac{1}{\hbar}\sum_{\begin{smallmatrix}d\geq 0 \\ \mu\end{smallmatrix}}
(\s_{d+2})_{\un,\mu}t_{\mu,d} 
\\ \notag
& +\frac{1}{2\hbar}\sum_{\begin{smallmatrix}i,j\geq 0 \\ \mu,\nu \end{smallmatrix}}
(-1)^{i}(\s_{i+j+1})_{\mu,\nu}t_{\mu,i}t_{\nu,j}
\\ \notag
& -\sum_\mu(\s_1)^{\mu}_{\un}\frac{\partial}{\partial t_{\mu,0}}
+\sum_{\begin{smallmatrix} d\geq 0,\ell\geq 1 \\ \mu,\nu \end{smallmatrix}}
(\s_\ell)^{\mu}_{\nu}t_{\nu,d+\ell}\frac{\partial}{\partial t_{\mu,d}}.
\end{align}

The geometrical meaning of the actions of these two parts of the Givental group is quite different: the upper triangular part corresponding to $\mathfrak{g}_+$ deforms the structure of the underlying Frobenius manifold, whereas the lower triangular part doesn't: it only changes the calibration of the Frobenius manifold as well as shifts the point around which one expands the potential $F$, see~\cite{FabShaZvo}.

\subsubsection{Convention on signs} We have fixed from the very beginning that we work with the metric given by the unit matrix in our choice of coordinates. In order to make clear the signs in the formula for $\hat{\r}$ and $\hat{\s}$, we use the following conventions for the shifts of indices: 
\begin{align*}
(\r_\ell)_\alpha^\beta& =(\r_\ell)^{\alpha\beta}, &(\s_\ell)_\alpha^\beta& =(\s_\ell)^{\alpha\beta}\\
(\r_\ell)_{\alpha\beta}& =(-1)^{\ell+1}(\r_\ell)^\alpha_\beta,&(\s_\ell)_{\alpha\beta}& =(-1)^{\ell+1}(\s_\ell)^\alpha_\beta.
\end{align*}
We use this convention throughout the rest of the paper, in all computations.


\section{The hierarchy associated to a potential}\label{sec:hierarchy}

In this section we describe the integrable hierarchy associated to a formal partition function. At first we shall be concerned with a formal neighbourhood of a non-homogeneous Frobenius manifold (that is, a Frobenius manifold without an Euler vector field). This is given by a tame $g=0$ potential $F_0$, i.e., satisfying the genus $0$ topological recursion relation (equivalent to the Equation~\eqref{trr} given in Definition~\ref{tame}) and the string equation~\eqref{string-equation}.

\subsection{The principal hierarchy}

\subsubsection{Notations for the calculus of variations}
The principal hierarchy associated to a Frobenius manifold is a system of partial differential equations in the variational bicomplex of functionals on the formal loop space of maps from $S^1$ to $V$. Explicitly, this means that if we denote the global coordinate on $S^1$ by $x$ and let $v_\alpha,~\alpha=1,\ldots, s$ be a basis of $V$, a formal loop in $V$ is parametrized by the jet coordinates $v_{\alpha,k}:=\partial^k v_\alpha/\partial x^k$, $k\geq 0$. On this formal loop space, we consider local functionals of the form
\begin{equation}
F(v):=\int f(x,v_0,\ldots, v_k)dx,
\end{equation}
where $f(x,v_0,\ldots, v_k)$ is a differential polynomial, i.e., depends analytically on $x$ and $v_{\alpha,0}$ and is a polynomial in the higher variables $v_{\alpha,k}$, $k\geq 1$.
The total derivative acting on such differential polynomials is given by
\begin{equation}
\partial_xf:=\frac{\partial f}{\partial x}+\sum_{\alpha,k}\frac{\partial f}{\partial v_{\alpha,k}}v_{\alpha, k+1},
\end{equation}
so that $\int \partial_x f dx=0$. Remark that with this definition one indeed has that $v_{\alpha,k+1}=\partial_x v_{\alpha,k}$. As a functional, the variational derivative is defined as 
\begin{equation}
\frac{\delta f}{\delta v_\alpha}:=\sum_{s=0}^\infty(-1)^s\partial_x^s\frac{\partial f}{\partial v_{\alpha,s}}.
\end{equation}
(Here we abuse a little bit the standard notations, where one used to write the variational derivative above applied to functionals $\int f \,dx$.)
For a detailed account of the variational bicomplex associated to the formal loop space, one should consult~\cite[\S 2.2.]{DubZha2}. 

For the construction of the principal hierarchy, we shall use the coordinates $v_\alpha$ defined in~\eqref{coord}. First we introduce an $x$-dependence by shifting along the $t_{\un,0}$, and define
\begin{equation}
\label{coord-x}
v_\alpha(x,t):=\frac{\partial^2 F_0}{\partial t_{\alpha,0}\partial t_{\un,0}}(x+t_{\un,0},t_1,t_2,\dots).
\end{equation}
With this shift we clearly have $\partial_xv_\alpha(x,t)=\partial v_\alpha(x,t)\slash\partial t_{\un,0}$ and therefore
\begin{equation}
v_{\alpha,k}(x,t)=\frac{\partial^{k+2} F_0}{\partial t_{\un,0}^{k+1}\partial t_{\alpha,0}}(t_{\un,0}+x,t_1,t_2,\dots).
\end{equation}

\subsubsection{The equations of the hierarchy}
Clearly, $v_\alpha(x,t)$ is a solution of the system of equations
\begin{equation}
\label{principal-hierarchy}
\frac{\partial v_\alpha}{\partial t_{\beta,q}}=\partial_x\left(\Omega^{[0]}_{\alpha,0,\beta,q}(v,0,0\dots)\right),\qquad \beta=1,\dots,s,\quad q\geq 0,
\end{equation}
since the left and right hand sides are equal to the same triple derivative of $F_0$. This system of equations is called the \emph{principal hierarchy} associated to the Frobenius manifold. More specifically, if we deal with a conformal Frobenius structure, i.~e., if we have an Euler vector field, this system of equations is a dispersionless bi-Hamiltonian hierarchy with $\tau$-symmetry. 

Without an Euler vector field, we have only one Hamiltonian structure that we are going to describe. We first introduce the following Poisson bracket on the formal loop space:
\begin{equation}
\label{poisson-principal}
\{F,G\}:=\int \sum_\alpha \frac{\delta f}{\delta v_\alpha}\partial_x\frac{\delta g}{\delta v_\alpha} dx,
\end{equation}
where $f$ and $g$ are the polynomial densities of the functionals $F$ and $G$.
Next we define the densities of higher Hamiltonians $H_{\alpha,p}$, $\alpha=1,\dots,s$, $p\geq 0$:
\begin{equation}
\label{ham-principal}
h_{\alpha,p}(v):=\Omega_{\un,0;\alpha,p+1}(v,0,0\dots).
\end{equation}

With respect to the Poisson bracket above, we have the following
\begin{proposition}
The Hamiltonians $H_{\alpha,p}=\int h_{\alpha,p}dx$ Poisson commute: 
\[
\{H_{\alpha,p},H_{\beta,q}\}=0.
\]
\end{proposition}
\begin{proof}
We need to show that 
\begin{equation}
\sum_\gamma \frac{\delta h_{\alpha,p}}{\delta v_\gamma}\partial_x\frac{\delta h_{\beta,q}}{\delta v_\gamma}
\end{equation}
is $\d_x$-exact. In fact, we can prove that this expression is equal to $\partial_x\Omega^{[0]}_{\alpha,p+1;\beta,q}$.
This is a straighforward computation using the topological recursion relations~\eqref{trr}, which we write in the coordinates $v_\gamma$ as
\begin{equation}\label{eq:trr-omega}
\frac{\partial\Omega^{[0]}_{\alpha,p+1;\beta,q}}{\partial v_\gamma}=\sum_\xi\Omega^{[0]}_{\alpha,p;\xi,0}\frac{\partial\Omega^{[0]}_{\xi,0;\beta,q}}{\partial v_\gamma}
\end{equation}
With this, we simply write out the Poisson bracket:
\begin{align}
\sum_\gamma \frac{\delta h_{\alpha,p}}{\delta v_\gamma}\partial_x\frac{\delta h_{\beta,q}}{\delta v_\gamma}
&=\sum_\gamma\frac{\partial\Omega^{[0]}_{\alpha,p+1;\un,0}}{\partial v_\gamma}\partial_x
\left(\frac{\partial\Omega^{[0]}_{\beta,q+1;\un,0}}{\partial v_\gamma}\right)
\\ \notag
&=\sum_{\gamma,\xi}\Omega^{[0]}_{\alpha,p;\xi,0}\frac{\partial\Omega^{[0]}_{\xi,0;\un,0}}{\partial v_\gamma}\partial_x\left(\frac{\partial\Omega^{[0]}_{\beta,q+1;\un,0}}{\partial v_\gamma}\right)
\\ \notag
&=\sum_{\gamma,\xi}\Omega^{[0]}_{\alpha,p;\gamma,0}
\partial_x\left(
\Omega^{[0]}_{\beta,q;\xi,0}\frac{\partial\Omega^{[0]}_{\xi,0;\un,0}}{\partial v_\gamma}
\right)
\\ \notag
&=\sum_\gamma\Omega^{[0]}_{\alpha,p;\gamma,0}\partial_x\Omega^{[0]}_{\gamma,0,\beta,q}
\\ \notag
&=\partial_x\Omega^{[0]}_{\alpha,p+1;\beta,q},
\end{align}
where in the last step one uses the fact that $\partial_x=\partial\slash\partial t_{\un,0}$ because of the shift of variables, together with the topological recursion relation~\eqref{eq:trr-omega} once again. This completes the proof.
\end{proof}

Combining this proposition with the hierarchy \eqref{principal-hierarchy} we find
\begin{equation}
\{v_\alpha,H_{\beta,q}\}=\partial_x\Omega^{[0]}_{\alpha,0,\beta,q}=\frac{\partial v_\alpha}{\partial t_{\beta,q}},
\end{equation}
meaning that the Hamiltonian vector field associated to $H_{\beta,q}$ is given by $\partial\slash\partial t_{\beta,q}$. 

In the presence of an Euler vector field, and, therefore, the second Hamiltonian structure, it is proved in~\cite{DubZha2} that this set of Hamiltonians is complete, justifying the name integrable hierarchy. The solution $v_{\alpha}(x,t)$ in~\eqref{coord-x} of equation \eqref{principal-hierarchy} is called the \emph{topological solution}. Other solutions can be constructed using the hodographic method, cf.~\cite[\S 3.7.4]{DubZha2}: they are called \emph{monotone solutions} and determined by an invertible element $u_{\alpha,1}(0)\in V$.


\subsection{The full hierarchy}

\subsubsection{Change of coordinates}
For the full, i.e., dispersive hierarchy, we consider the formal extended loop space, meaning that we now consider formal series
\begin{equation}
f(x,w_0,w_1,\dots,\hbar)=\sum_{k=0}^\infty \hbar^k f_k(x,w_{0},w_{1},\dots, w_{2k}),
\end{equation}
where each $f_k$ is a differential polynomial in $w_1,\dots,w_{2k}$ of degree $2k$, $\deg(w_{\alpha,i})=i$.\footnote{This differs a little bit from the original conventions of Dubrovin and Zhang. They consider series in $\epsilon=\sqrt{\hbar}$, and the coefficient of $\epsilon^k$ is a weighted homogeneous polynomial of degree $k$.} The natural group of coordinate transformations on the extended loop space is the so-called \emph{Miura group} of formal diffeomorphisms
\begin{equation}
w_\alpha\mapsto \tilde{w}_\alpha:=\sum_{k=0}^\infty\hbar^k G_{\alpha,k}(w,w_{1},\dots,w_{2k}),
\end{equation}
where $G_{\alpha,0}$, $\alpha=1,\dots,s$ is an invertible coordinate transformation and $G_{\alpha,k}$ are differential polynomials in $w_1,\dots,w_{2k}$ with $\deg(G_{\alpha,k})=2k$. When each $G_{\alpha,k}$ is a rational function of degree $2k$ (and, therefore, $G_{\alpha,k}$ might depend on higher derivatives than $w_{2k}$, but still on a finite number of them), such a coordinate change is called a \emph{quasi-Miura} transformation.

We now consider the full partition function \eqref{potential}, and introduce the coordinates
\begin{equation}
w_\alpha(t_0,t_1,\dots):=\frac{\partial^2 \left(\sum_{g=0}^\infty \hbar^g F_g\right)}{\partial t_{\alpha,0}\partial t_{\un,0}}(t_0,t_1,t_2,\dots).
\end{equation}
Again we introduce the $x$-variable by shifting along the $t_{\un,0}$-direction: $w_\alpha(x,t):=w_\alpha(t_{0}+x,t_1,t_2,\dots)$, and therefore
\begin{equation}
w_{\alpha,k}=\frac{\partial^{k+2}\left(\sum_{g=0}^\infty \hbar^g F_g\right)}{\partial^{k+1}t_{\un,0}\partial t_{\alpha,0}}(t_0+x,t_1,t_2,\dots).
\end{equation}

Recall that $F_g$ is a function of $3g-1$ variables $v,v_1,\dots,v_{3g-2}$ as given by Equation~\eqref{eq:3g-2}. Therefore, the second derivative of $F_g$ depends on $v,v_1,\dots,v_{3g}$ (here we have to use the principal hierarchy in order to turn the derivatives in $t$-variables into the derivatives in $x$-variables). So, the change of variables that we have here looks like
\begin{equation}\label{eq:v-w}
v_\alpha\mapsto {w}_\alpha:=v_\alpha+\sum_{g=1}^\infty \hbar^g \frac{\d^2 F_g}{\d x\d t_{\alpha,0}} (v,v_{1},\dots,v_{3g}),
\end{equation}

In the case of a conformal Frobenius structure, Dubrovin and Zhang prove in~\cite{DubZha2} that it is a quasi-Miura transformation. In general case, we have no control on how bad are the coefficients of the $\hbar$-expansion of this change of variables, though they still depend on a finite number of the derivatives of the coordinates $v_\alpha$. We call such changes of variables \emph{weak quasi-Miura transformations}.

\subsubsection{Ingredients of the full hierarchy} \label{sec:ingr}
Following the genus zero theory, we now define
\begin{equation}
\Omega_{\alpha,p,\beta,q}:=\sum_{g=0}^\infty \hbar^g \frac{\partial^2 F_g}{\partial t_{\alpha,p}\partial t_{\beta,q}}.
\end{equation}
We see that $w_{\alpha}(x,t)$ is a solution of the system of partial differential equations
\begin{equation}
\label{full-hierarchy}
\frac{\partial w_\alpha}{\partial t_{\beta,q}}=\partial_x\left(\Omega_{\alpha,0,\beta,q}(w,w_1,w_2,\dots)\right).
\end{equation}

This system of partial differential equations is again a Hamiltonian system obtained from the principle hierarchy by the weak quasi-Miura transformation~\eqref{eq:v-w}. This coordinate change transforms the Poisson bracket~\eqref{poisson-principal} to another Poisson bracket given by the formula
\begin{equation}
\{F,G\}:=\int \sum_{\alpha,\beta} \frac{\delta f}{\delta w_\alpha}\sum_{s=0}^\infty A^{\alpha\beta}_{s}\d_x^s\frac{\delta g}{\delta w_\beta} dx,
\end{equation}
where $A_s^{\alpha\beta}=\sum_{g=0}^\infty \hbar^g A_{g,s}^{\alpha\beta}$ is a formal power series in $\hbar$ whose coefficients are some functions in $w,w_1,w_2,\dots$ given by the formula
\begin{equation}
\sum_{s=0}^\infty A_s^{\alpha\beta}\d_x^s := \sum_{\begin{smallmatrix} \mu,e \\ \nu,f \end{smallmatrix}} 
\frac{\d w_\alpha}{\d v_{\mu,e}} \d_x^e \circ \d_x\circ (-\d_x)^f \circ \frac{\d w_\beta}{\d v_{\nu,f}}.
\end{equation}
Is is not immediately obvious, but it is very easy to show (see Section~\ref{sec:unique} below) that $A_0^{\alpha\beta}=0$. 

Since we have no control on weak quasi-Miura transformations, we can't say anything about what kind of function $A_{g,s}^{\alpha\beta}$ is. In the case of a conformal Frobenius structure, when the coordinate change is a quasi-Miura transformation, Dubrovin and Zhang conjecture that it is a homogeneous polynomial in $w_1,w_2,\dots$ of degree $2g-s$ (we assume, as usual, that $\deg w_i=i$).
We prove this conjecture for an arbitrary semi-simple Frobenius structure in Section~\ref{sec:conclusions}.

In principle, under the coordinate change the Hamiltonians of the full hierarchy should be simply recalculated in the new coordinates. However, there is still a freedom for the choice of densities of the Hamiltonians, since we can always add a $\d_x$-exact term to them. It is, therefore, natural to define the densities of the Hamiltonians equal to
\begin{align}
\label{full-ham}
h_{\alpha,p}(w)& :=\Omega_{\alpha,p+1;\un,0}(w,w_1,w_2\dots) 
\\ \notag
& =\sum_{g=0}^\infty \hbar^{g} \frac{\partial^2 F_g}{\partial t_{\un,0}\partial t_{\alpha,p+1}}(w,w_1,w_2,\dots),
\end{align}
which is simply the densities of the Hamiltonians~\eqref{ham-principal} deformed by $\d_x(\sum_{g\geq 1} \hbar^g \d F_g/\d t_{\alpha,p+1})$.

In the case of a conformal Frobenius structure, when the coordinate change is a quasi-Miura transformation, Dubrovin and Zhang conjectured and even attempted to prove that in the variables $w_1,w_2,\dots$ the coefficient of $\hbar^g$ of the $\hbar$-expansion of any $\Omega_{\alpha,p;\beta,q}$ is a homogeneous polynomial of degree $2g$. As we have already mentioned above, unfortunately, Boris Dubrovin has informed us that they have found a gap in their argument. We generalize their conjecture for an arbitrary semi-simple Frobenius structure and prove it in Section~\ref{sec:conclusions}.

\begin{example}[The KdV hierarchy] \label{KdV}
The fundamental example of a principal and full hierarchy associated to a tame partition function is given by the KdV hierarchy. It is associated to the Gromov-Witten potential of the point, or, simple, the generating function of the intersection number of $\psi$-classes on the moduli space of curve,
\begin{equation}
Z_{KdV}:=\exp\left(\sum_{g=0}^\infty \hbar^{g-1}\sum_{
\begin{smallmatrix}
n\geq 1 \\ 2g-2+n>0
\end{smallmatrix}
} \frac{1}{n!} \sum_{
d_1,\dots,d_n\geq 0
}
\int_{\overline{\mathcal{M}}_{g,n}}\prod_{i=1}^n\psi_i^{d_i}t_{d_i}
\right).
\end{equation}
This corresponds to a one-dimensional Frobenius manifold, that is, $\dim(V)=1$, with prepotential $F_0(v)=v^3/6$. The hierarchy can be given very conveniently in Lax form. Writing out, the first few equations read
\begin{align}
\label{kdv}
w_{t_0}&=w_x,\\ \notag
w_{t_1}&=ww_x+\frac{\hbar}{12}w_{xxx}\\ \notag
w_{t_2}&=\frac{1}{2}w^2w_x+\frac{\hbar}{12}(2w_xw_{xx}+ww_{xxx})+\frac{\hbar^2}{240}w_{xxxxx}\\ \notag
& \vdots
\end{align}
What is important for us, is that it is an example of a bi-Hamiltonian hierarchy, as shown in \cite{DubZha2}, 
with the first Poisson bracket given by~\eqref{poisson-principal} and the Hamiltonians given by
\begin{align}
h_{-1}&=w\\ \notag
h_0&=\frac{w^2}{2}+\hbar\frac{w_{xx}}{12}\\ \notag
h_1&=\frac{w^3}{6}+\frac{\hbar}{24}(w_x^2+2ww_{xx})+\hbar^2\frac{w_{xxxx}}{240}\\ \notag
& \vdots
\end{align}

Setting $\hbar=0$ one finds the dispersionless limit of KdV, also called the Riemann hierarchy. It is proved in 
 \cite[\S 3.8.3]{DubZha2} that the transformation from the Riemann hierarchy to the full KdV hierarchy given by
 \begin{equation}
 v\mapsto v+\frac{\hbar}{24}(\log v_x)_{xx}+\hbar^2\left(\frac{v_{xxxx}}{1152 v_x^2}-\frac{7v_{xx}v_{xxx}}{1920 v_x^3}+\frac{v_{xx}^3}{360 v_x^4}\right)_{xx}+\mathcal{O}(\hbar^3),
 \end{equation}
is a quasi-Miura transformation.

It is a special feature of the KdV hierarchy that its Poisson bracket remains undeformed when going from the dispersionless hierarchy to the dispersive tail. It shows explicitly that the Poisson bracket is polynomial, so we can use it in our argument as the basepoint under the action of the Givental group.
\end{example}


\section{Deformation formulas for $\Omega_{\alpha,p;\beta,q}$} \label{sec:def-omega}

In this section, we obtain formulas for the infinitesimal deformations of $\Omega_{\alpha,p;\beta,q}$ as a function of $w_{\gamma,n}$ (defined in Section~\ref{sec:ingr}) by elements of the Lie algebra of the Givental group (presented in Section~\ref{sec:Givental}). We write $\r_\ell z^\ell\in\mathfrak{g}_+$ and $\s_\ell z^\ell\in\mathfrak{g}_-$ for generic elements in the Lie algebra of the Givental group. Their action on a (multiple derivative of a)  tame partition function is denoted by a lower dot.  In \eqref{eq:r-action} and \eqref{eq:s-action}, this action is given in terms of the $t$-variables. When we consider the resulting function in other coordinates, in this case $w_{\gamma,n}$, we write this coordinate in square brackets behind the Lie algebra element.
\begin{theorem} \label{thm:r-def-Omega} We have:
\begin{align} \label{eq:def-Omega}
& \widehat{\r_\ell z^\ell}[w]. \Omega _{\alpha,p;\beta,q} = 
 (\r_\ell)_\alpha^\mu \Omega_{\mu,p+\ell;\beta,q} + \Omega_{\alpha,p;\mu,q+\ell}(\r_\ell)_\beta^\mu \\ \notag
& \phantom{ =\ } +\sum_{i=0}^{\ell-1} (-1)^{i+1} \Omega_{\alpha,p;\mu,i} (\r_\ell)^{\mu\nu} \Omega_{\nu,\ell-1-i;\beta,q} \\ \notag
& \phantom{ =\ } - \sum_{\gamma,n} \frac{\d \Omega _{\alpha,p;\beta,q}}{\d w_{\gamma,n}} \left(  
(\r_\ell)^\mu_\gamma \d_x^n \Omega_{\mu,\ell;\un,0} 
+ (n+1)\d_x^n \Omega _{\gamma,0;\mu,\ell} (\r_\ell)^\mu_\un \phantom{\sum_{i=1}^{\ell-1}}
\right. \\ \notag
& \phantom{ = -\ }
+\sum_{i=0}^{\ell-1}\sum_{k=0}^{n-1} \binom{n}{k} (-1)^{i+1} \d_x^{k+1}\Omega_{\gamma,0;\mu,i} (\r_\ell)^{\mu\nu} \d_x^{n-k-1}\Omega_{\nu,\ell-1-i;\un,0} \\
& \phantom{ = -\ } \notag
\left.
+\sum_{i=0}^{\ell-1} (-1)^{i+1} \d_x^n \left(\Omega_{\gamma,0;\mu,i} (\r_\ell)^{\mu\nu} \Omega_{\nu,\ell-1-i;\un,0} \right)
\right) \\ \notag
& 
+\frac{\hbar}{2} \sum_{
\begin{smallmatrix}
\gamma, n\\
\zeta, m
\end{smallmatrix}} \frac{\d^2 \Omega _{\alpha,p;\beta,q}}{\d w_{\gamma,n}\d w_{\zeta,m}}
\sum_{i=0}^{\ell-1}(-1)^{i+1}
\d_x^{n+1} \Omega_{\gamma,0;\mu,i} (\r_\ell)^{\mu\nu} \d_x^{m+1} \Omega_{\nu,\ell-1-i;\zeta,0}.
\end{align}
\end{theorem}

\begin{proof} Direct computation. One should just use the formula 
\begin{equation}
\widehat{\r_\ell z^\ell}[w]. \Omega _{\alpha,p;\beta,q} = \widehat{\r_\ell z^\ell}[t]. \Omega _{\alpha,p;\beta,q} - \sum_{\gamma,n}\frac{\d \Omega _{\alpha,p;\beta,q}}{\d w_{\gamma,n}} \cdot \d_x^n \widehat{\r_\ell z^\ell}[t]. \Omega _{\gamma,0;\un,0},
\end{equation}
which is the change of coordinates from $t$ to $w$. 
Here $\widehat{\r_\ell z^\ell}[t]. \Omega _{\alpha,p;\beta,q}$ (and $\widehat{\r_\ell z^\ell}[t]. \Omega _{\gamma,0;\un,0}$) is just 
\begin{equation}
\frac{\d^2}{\d t_{\alpha,p}\d t_{\beta,q}} \sum_{g=0}^\infty \hbar^g \widehat{\r_\ell z^\ell}[t].F_g(t),
\end{equation}
where $\widehat{\r_\ell z^\ell}[t].F_g(t)$ is given by the formulas of Y.-P.~Lee~\cite{Lee1,FabShaZvo}, cf. \eqref{eq:r-action} and \eqref{eq:s-action}.
\end{proof}

\begin{remark} We can simplify the formula~\eqref{eq:def-Omega}. We introduce a new notation. If $p<0$ or $q<0$, we set $\Omega_{\alpha,p;\beta,q}$ to be equal to $(-1)^p\delta_{\alpha\beta}\delta_{p+q,-1}$ if $p$ is nonnegative and to $(-1)^q\delta_{\alpha\beta}\delta_{p+q,-1}$ if $q$ is nonnegative. Then we can rewrite equation~\eqref{eq:def-Omega} as
\begin{align}\label{eq:def-Omega mod}
& \widehat{\r_\ell z^\ell}[w]. \Omega _{\alpha,p;\beta,q} = \sum_{d=-\infty}^{\infty} (-1)^{d+1} (\r_\ell)^{\mu\nu} \left[ \Omega_{\alpha,p;\mu,d} \Omega_{\nu,\ell-1-d;\beta,q} \phantom{\sum_{
\zeta, m
}} \right.\\ \notag
& \phantom{ =\ } - \sum_{\gamma,n} \frac{\d \Omega _{\alpha,p;\beta,q}}{\d w_{\gamma,n}} 
\sum_{a=0}^{n} \binom{n+1}{a} \d_x^{a}\Omega_{\gamma,0;\mu,d} \d_x^{n-a}\Omega_{\nu,\ell-1-d;\un,0} \\ \notag
& \phantom{ =\ } 
+\frac{\hbar}{2} \sum_{
\begin{smallmatrix}
\gamma, n\\
\zeta, m
\end{smallmatrix}} \left. \frac{\d^2 \Omega _{\alpha,p;\beta,q}}{\d w_{\gamma,n}\d w_{\zeta,m}}
\d_x^{n+1} \Omega_{\gamma,0;\mu,d} \d_x^{m+1} \Omega_{\nu,\ell-1-d;\zeta,0}\right].
\end{align}
\end{remark}

We obtain a similar formula for the $\s$-action.

\begin{theorem}
We have:
\begin{align} \label{eq:def-Omega-low}
& \sum_{\ell=1}^\infty \widehat{\s_\ell z^\ell}[w]. \Omega _{\alpha,p;\beta,q} = 
\sum_{1\leq\ell\leq p}(\s_\ell)^{\mu}_{\alpha}\Omega_{\mu,p-\ell;\beta,q}+\sum_{1\leq\ell\leq q}
\Omega_{\alpha,p;\mu,q-\ell} (\s_\ell)^{\mu}_{\beta} 
\\ \notag
& + (-1)^p(\s_{p+q+1})_{\alpha,\beta}-\sum_{\gamma}\frac{\partial\Omega_{\alpha,p;\beta,q}}{\partial w_{\gamma,0}} (\s_1)_{\gamma,\un}.
\end{align}
\end{theorem}

\begin{proof} The proof is again a straighforward computation of the same kind as in the proof of Theorem~\ref{thm:r-def-Omega}.
\end{proof}

\begin{remark}
We can rewrite equation~\eqref{eq:def-Omega-low} as
\begin{align}
& \sum_{\ell=1}^\infty \widehat{\s_\ell z^\ell}[w]. \Omega _{\alpha,p;\beta,q} = \sum_{\ell=1}^{\infty}\sum_{i+j=-l-1}(-1)^{i+1}\Omega_{\alpha,p;\mu,i}(\s_\ell)^{\mu\nu}\Omega_{\nu,j;\beta,q}\\
& -\sum_{\gamma}\frac{\partial\Omega_{\alpha,p;\beta,q}}{\partial w_{\gamma,0}} (\s_1)_{\gamma,\un}.\notag
\end{align}
\end{remark}

\begin{remark}
In genus $0$, the functions $\Omega^{[0]}_{\alpha,p;\beta,q}$ form a symmetric solution of a so-called \emph{master equation}~\cite{ShaZvo}, which is an extension of commutativity equations~\cite{LosPol1,LosPol2,LosMan}. There is a Givental-type theory of deformations of solutions of commutativity equations developed in~\cite{LosPol2} and revisited in~\cite{ShaZvo}. The deformation formulas there are given simply by the first three summands in Equations~\eqref{eq:def-Omega} and~\eqref{eq:def-Omega-low}, and they have a very nice interpretation in terms of multi-KP hierarchies~\cite{Leu,FeiLeuSha}, geometry of the Losev-Manin moduli spaces~\cite{LosMan}, and Givental-type linear algebra of the loop space, see~\cite{ShaZvo} for a detailed discussion.
\end{remark}


\section{Deformation formulas for a bracket} \label{sec:def-bracket}

In this section we obtain a deformation formula for a Poission bracket that gives one of the two Poisson structures for the Dubrovin-Zhang hierarchies. The starting point for this calculation is the equations of the hierarchy, written out using the Poisson bracket:
\begin{equation}
\{w_\beta,h_{\alpha,p}\}=\partial_x\Omega_{\alpha,p;\beta,0}.
\end{equation}
Using the deformation formulas for the $\Omega_{\alpha,p;\beta,q}$ of the previous section, we obtain deformations of the densities of Hamiltonians $h_{\alpha,p}$ and $w_\beta$, as well as the right hand side of the equation above. In the case of the $\r$-action, we are therefore looking for a differential operator $\sum_{s=1}^\infty\left(\widehat{\r_\ell z^\ell}[w]. A_{s}^{\alpha\beta}\right) \d_x^s$ such that 
\begin{align} \label{eq:def-A}
\d_x \widehat{\r_\ell z^\ell}[w]. \Omega_{\alpha,p;\beta,0}
& = \sum_{\gamma}\sum_{s=1}^\infty\left(\widehat{\r_\ell z^\ell}[w]. A_{s}^{\beta\gamma}\right) \d_x^s \frac{\delta}{\delta w_\gamma}\Omega_{\alpha,p+1;\un,0} \\
& \phantom{ =\ } \notag
+ \sum_{\gamma}\sum_{s=1}^\infty A_{s}^{\beta\gamma} \d_x^s \frac{\delta}{\delta w_\gamma}\widehat{\r_\ell z^\ell}[w].\Omega_{\alpha,p+1;\un,0}.
\end{align}

Equation~\eqref{eq:def-A} has a quite involved solution, so we first need to introduce some new notations. In Section~\ref{sec:unique} we discuss the uniqueness of this solution.

\subsection{Some notations} In order to shorten some intermediate formulas, we introduce the following notations:
\begin{align}
\delta_{\xi}& := \frac{\delta}{\delta w_\xi} = \sum_{n=0}^\infty (-\d_x)^n\circ \frac{\d}{\d w_{\xi,n}};
& \d_{\xi,n} & := \frac{\d}{\d w_{\xi,n}}; \\ \notag
\T_{\xi,k} & := \sum_{n=0}^\infty \binom{n}{k} (-\d_x)^{n-k} \circ\frac{\d}{\d w_{\xi,n}}.
\end{align}
We use the agreement that $\binom{n}{k}=0$ if $n\geq 0$ and $k<0$ or $k>n$. 

The operators $\T_{\xi,k}$ satisfy the following properties:
\begin{align}
\T_{\xi,0} & = \delta_\xi; &
\T_{\xi,k} & = 0 \mbox{ if } k<0; &
\T_{\xi,k} \circ \d_x & = \T_{\xi, k-1} \mbox{ for any } k\in\Z. 
\end{align}
Moreover, for any functions $X,Y$,
\begin{align}
\delta_\xi(XY) & =\sum_{k=0}^\infty \left( \T_{\xi,k}X  (-\d_x)^k Y  + (-\d_x)^k X  \T_{\xi,k} Y \right),
\end{align}
and, more generally, for any $p\geq 0$,
\begin{align}
\T_{\xi,p}(X Y) & =\sum_{k=0}^\infty \binom{k+p}{k} \left( \T_{\xi,k+p}X  (-\d_x)^k Y  + (-\d_x)^k X  \T_{\xi,k+p} Y \right)
\end{align}
(see~\cite{LiuZha} for more useful formulas of the same type).

Another notation that we are using is the following. We denote by $\Omega_{\gamma_1,k_1;\gamma_2,k_2;\gamma_3,k_3}$ the triple derivative $\d^3(\sum_{g=0}^\infty \hbar^g F_g)/\d t_{\gamma_1,k_1}\d t_{\gamma_2,k_2}\d t_{\gamma_3,k_3}$ considered as a function of $w_{\xi,n}$. It is a series in $\hbar$, and the coefficient at $\hbar^g$ is a weighted homogeneous polynomial in the derivatives $w_{\xi,n}$, $n\geq 1$, of degree $2g+1$. This follows from the following formula:
\begin{equation}
\Omega_{\gamma_1,k_1;\gamma_2,k_2;\gamma_3,k_3} := \sum_{\xi,n} \d_x^{n+1} \Omega_{\gamma_i,k_i;\xi,0} \frac{\d \Omega_{\gamma_j,k_j;\gamma_\ell,k_\ell}}{\d w_{\xi,n}}
\end{equation}
for any choice of indices $\{i,j,\ell\}=\{1,2,3\}$. We assume that $\Omega_{\gamma_1,k_1;\gamma_2,k_2;\gamma_3,k_3}$ is equal to $0$ if some of the indices $k_1,k_2,k_3$ is negative.
\begin{remark}
We make one final remark about the notation in which the deformation formula is presented below. In order to reduce the amount of brackets in the expressions, we write $\circ$ for the composition of differential operators. If a differential operator appears without composition to the right, it is to be applied to the expression immediate on the right of it. 
\end{remark}
\subsection{A formula for the operator of $\r$-deformation}

\begin{theorem}\label{thm:r-def-A} Equation~\eqref{eq:def-A} has the following solution: 
\begin{align}
& \sum_{s=1}^{\infty} \left(\widehat{\r_\ell z^\ell}[w]. A_{s}^{\beta\xi}\right) \d_x^s \\ \notag
& = \sum_{i+j=\ell-1} (-1)^{i+1} (\r_\ell)^{\mu\nu} 
\left[ 
\Omega_{\un,0;\nu,j} \sum_{\gamma,n} \frac{\d \Omega_{\mu,i;\beta,0}}{\d w_{\gamma,n}} \d_x^{n}\circ 
\sum_{s\ge 1,\xi}^\infty A^{\gamma,\xi}_s \d_x^s \right.
\\ 
\notag
& - \sum_{\gamma,n} \sum_{a+b=n} \binom{n+1}{a}\d_x^b \Omega_{\un,0;\nu,j} \d_x^a \Omega_{\mu,i;\gamma,0}  
\sum_{s\ge 1,\xi} \frac{\d A^{\beta,\xi}_s}{\d w_{\gamma,n}} \d_x^s  
\\ \notag
& + \sum_{s\ge 1,\gamma} A^{\beta,\gamma}_s \sum_{f+e=s-1} \d_x^f \circ \Omega_{\un,0;\nu,j}
\d_x^e \circ \sum_{n=0}^\infty \T_{\gamma,n}\Omega_{\mu,i;\xi,0} (-\d_x)^{n+1}
\\ 
\notag
& + \Omega_{\beta,0;\nu,j}\sum_{\gamma,n}\frac{\d \Omega_{\mu,i;\un,0}}{\d w_{\gamma,n}} \d_x^n\circ \sum_{s\ge 1,\xi} A^{\gamma,\xi}_s \d_x^s 
\\ \notag
& + \sum_{s\ge 1,\gamma} A^{\beta\gamma}_s \d_x^s\circ
\sum_{0\leq u\leq v} \binom{v}{u}
\T_{\gamma,v+1}\Omega_{\un,0;\nu,j} (-\d_x)^{v-u} \Omega_{\mu,i;\xi,0} (-\d_x)^{u+1}  
 \\ \notag
& - \sum_{s\ge 1,\gamma} A^{\beta\gamma}_s \d_x^s\circ \sum_{0\leq u\leq v} \binom{v+1}{u}
(-\d_x)^{v-u}\Omega_{\un,0;\nu,j} \T_{\gamma,v+1} \Omega_{\mu,i;\xi,0} (-\d_x)^{u+1}  
\\ \notag 
&  + \sum_{s\ge 1,\gamma} A^{\beta\gamma}_s \sum_{e+f=s-1} \binom{s}{e} \d_x^e 
\delta_\gamma \Omega_{\un,0;\nu,j} \d_x^f \circ \Omega_{\mu,i;\xi,0} \d_x 
\\ \notag &
-\d_x\Omega_{\beta,0;\nu,j-1}
\sum_{\gamma,m} \sum_{u=0}^{m-1} 
(-\d_x)^u \d_{\gamma,m} \Omega_{\mu,i+1;\un,0} \d_x^{m-1-u}\circ \sum_{s=1}^\infty A^{\gamma\xi}_s\d_x^s  
\\ \notag
& 
-\d_x\Omega_{\beta,0;\nu,j-1}  \sum_{\gamma}\sum_{2\leq f \leq s}^\infty 
(-\d_x)^{s-f} \left( A^{\gamma\xi}_s \delta_{\gamma} \Omega_{\mu,i+1;\un,0}\right) \d_x^{f-1}
\\ \notag & 
+\frac{\hbar}{2} \left( 
\d_x \circ 
\sum_{\gamma, n} \frac{\d \Omega_{\beta,0;\mu,i;\nu,j}}{\d w_{\gamma,n}}
\d_x^{n} 
\circ \sum_{s\ge 1}  A^{\gamma,\xi}_s \d_x^s \right.
\\ \notag
& +
\sum_{s\ge 1,\gamma}  A_{s}^{\beta\gamma} \d_x^s 
\circ 
\sum_{m=0}^\infty 
\T_{\gamma,m} \Omega_{\xi,0;\mu,i;\nu,j}
 (-\d_x)^{m+1}  \\ \notag
& \left.\left. - \sum_{n=0}^\infty\sum_{\zeta} \d_x^{n+1} 
\Omega_{\zeta,0;\mu,i;\nu,j} \sum_{s\ge 1} \frac{\d A_s^{\beta\xi}}{\d w_{\zeta,n}} 
\d_x^s\right)\right].
\end{align}
\end{theorem}

\begin{proof} The proof is based on an explicit computation of all terms in the formula~\eqref{eq:def-Omega mod} applied to $\Omega_{\alpha,p;\beta,0}$ and $\Omega_{\alpha,p+1;\un,0}$. This computation is performed in Sections~\ref{sec:usefull}-\ref{sec:hbar-0} below.\end{proof} 

\subsection{Useful lemmas}\label{sec:usefull}

There are two commutation relations that we are going to use several times. Consider an operator $\sum_{n=0}^\infty \d_x^n B \d_{\zeta,n}$, where $B$ is an arbitrary function. 

\begin{lemma}\label{lemma:d_x} We have: $[\d_x,\sum_{n=0}^\infty \d_x^n B \d_{\zeta,n}]=0$. 
\end{lemma}

\begin{proof} Since $[\d_x, \d_{\zeta,n+1}]=-\d_{\zeta,n}$, we have:
\begin{equation}
[\d_x, \sum_{n=0}^\infty \d_x^n B \d_{\zeta,n}] = \sum_{n=0}^\infty \d_x^{n+1} B \d_{\zeta,n}
-\sum_{n=1}^\infty \d_x^n B\d_{\zeta,n-1} = 0.
\end{equation}
\end{proof}

\begin{lemma}\label{lemma:commutator} For any function $A$, we have: 
\begin{align}
& [ A \d_x^s\circ{\delta_\gamma},\sum_{n=0}^\infty \d_x^n B \d_{\zeta,n}] \\ \notag
& = A \d_x^s \circ \sum_{j=0}^\infty \T_{\gamma,j} B (-\d_x)^j\circ {\delta_{\zeta}} 
- \sum_{n=0}^\infty \d_x^n B \frac{\d A}{\d w_{\zeta,n}}\d_x^s\circ {\delta_\gamma}
\end{align}
\end{lemma}

\begin{proof} Observe that
\begin{align}\label{eq:A-circ-B}
& A \d_x^s\circ {\delta_\gamma} \circ \sum_{n=0}^\infty \d_x^n B {\d_{\zeta,n}} \\ \notag
& = A \sum_{m,n=0}^\infty (-1)^m \sum_{i=0}^{s+m} \binom{s+m}{i}  \d_x^{n+i} B  \d_x^{s+m-i}\circ {\d_{\gamma,m}\circ\d_{\zeta,n}}\\ \notag
& + A\d_x^s\circ \sum_{j,n=0}^\infty\T_{\gamma,j}\d_x^n B (-\d_x)^j\circ {\d_{\zeta,n}}.
\end{align}
Since $\T_{\gamma,j}\circ\d_x^n=\T_{\gamma,j-n}$, the last summand is equal to 
\begin{align}\label{eq:higher}
& A\d_x^s\circ \sum_{j=0}^\infty \T_{\gamma,j} B (-\d_x)^j {\delta_{\zeta}}.
\end{align}
On the other hand,
\begin{align}\label{eq:B-circ-A}
& \sum_{n=0}^\infty \d_x^n B{\d_{\zeta,n}} \circ A \d_x^s\circ {\delta_\gamma} =
\sum_{n=0}^\infty  \d_x^n B  \frac{\d A}{\d w_{\zeta,n}}\d_x^s\circ {\delta_\gamma} \\ \notag
& + \sum_{m,n=0}^\infty (-1)^m \sum_{i=0}^{s+m} \binom{s+m}{i}  \d_x^{n+i} B A \d_x^{s+m-i}\circ {\d_{\gamma,m}\circ \d_{\zeta,n}}.
\end{align}
We see that the difference of the expressions in~\eqref{eq:A-circ-B} and~\eqref{eq:B-circ-A} gives exactly the statement of the lemma.
\end{proof}

\subsection{The coefficient of $\hbar^1$}
First, let us rewrite the $\hbar$-term on the righthand side of equation~\eqref{eq:def-Omega mod}. Let $i+j=l-1$, we have:
\begin{align}
& \sum_{
\begin{smallmatrix}
\gamma, n\\
\zeta, m
\end{smallmatrix}} \frac{\d^2 \Omega _{\alpha,p;\beta,q}}{\d w_{\gamma,n}\d w_{\zeta,m}}
\d_x^{n+1} \Omega_{\gamma,0;\mu,i} \d_x^{m+1} \Omega_{\nu,j;\zeta,0} \\ \notag
& = \left[ \sum_{\gamma, n} \d_x^{n+1} \Omega_{\gamma,0;\mu,i} {\d_{\gamma,n}}
\circ 
\sum_{\zeta, m} \d_x^{m+1} \Omega_{\nu,j;\zeta,0} {\d_{\zeta,m}}
 \right. \\ \notag
& \left. 
- \sum_{\zeta, m} \d_x^{m+1} \Omega_{\mu,i;\nu,j;\zeta,0} {\d_{\zeta,m}}
\right] \Omega_{\alpha,p;\beta,q}
\end{align}
(here we used the formula $\frac{\d}{\d t_{\delta,r}}=\sum_{\gamma,n}\d_x^{n+1}\Omega_{\delta,r;\gamma,0}\d_{\gamma,n}$).
Observe that since ${\delta_\xi}\circ\d_x =0$, we have:
\begin{align}
& {\delta_\xi}  
\sum_{\gamma, n} \d_x^{n+1} \Omega_{\gamma,0;\mu,i} {\d_{\gamma,n}}
\sum_{\zeta, m} \d_x^{m+1} \Omega_{\nu,j;\zeta,0} {\d_{\zeta,m}} \Omega_{\alpha,p+1;\un,0} \\ \notag
& = {\delta_\xi} \d_x \Omega_{\alpha,p+1;\mu,i;\nu,j} 
 = 0.
\end{align}
Also observe that
\begin{align}
&
\sum_{\gamma, n} \d_x^{n+1} \Omega_{\gamma,0;\mu,i} {\d_{\gamma,n}}
\sum_{\zeta, m} \d_x^{m+1} \Omega_{\nu,j;\zeta,0} {\d_{\zeta,m}} \Omega_{\alpha,p;\beta,0} \\ \notag
& =
\sum_{\gamma, n} \frac{\d \Omega_{\beta,0;\mu,i;\nu,j}}{\d w_{\gamma,n}}
\d_x^{n}
\sum_{s\ge 1,\xi} A^{\gamma,\xi}_s \d_x^s {\delta_{\xi}} \Omega_{\alpha,p+1;\un,0}.
\end{align}
Using these observations, Lemma~\ref{lemma:d_x} and Lemma~\ref{lemma:commutator}, we obtain the following expression: 
\begin{align}
& \d_x \sum_{
\begin{smallmatrix}
\gamma, n\\
\zeta, m
\end{smallmatrix}} \frac{\d^2 \Omega _{\alpha,p;\beta,0}}{\d w_{\gamma,n}\d w_{\zeta,m}}
\d_x^{n+1} \Omega_{\gamma,0;\mu,i} \d_x^{m+1} \Omega_{\nu,j;\zeta,0} \\ \notag
& - \sum_{s\ge 1,\xi}  A_{s}^{\beta\xi} \d_x^s {\delta_\xi}  \sum_{
\begin{smallmatrix}
\gamma, n\\
\zeta, m
\end{smallmatrix}} \frac{\d^2 \Omega _{\alpha,p+1;\un,0}}{\d w_{\gamma,n}\d w_{\zeta,m}} 
\d_x^{n+1} \Omega_{\gamma,0;\mu,i} \d_x^{m+1} \Omega_{\nu,j;\zeta,0}  \\ \notag
& = \left[ 
\d_x \circ 
\sum_{\gamma, n} \frac{\d \Omega_{\beta,0;\mu,i;\nu,j}}{\d w_{\gamma,n}}
\d_x^{n} 
\circ \sum_{s\ge 1,\xi}  A^{\gamma,\xi}_s \d_x^s \circ {\delta_{\xi}} \right.
\\ \notag
& +
\sum_{s\ge 1,\xi}  A_{s}^{\beta\xi} \d_x^s 
\circ 
\sum_{m=0}^\infty\sum_{\zeta} 
\T_{\xi,m} \d_x\Omega_{\zeta,0;\mu,i;\nu,j}
 (-\d_x)^{m} \circ {\delta_{\zeta}}  \\ \notag
& \left. - \sum_{n=0}^\infty\sum_{\zeta} \d_x^{n+1} 
\Omega_{\zeta,0;\mu,i;\nu,j} \sum_{s\ge 1,\xi} \frac{\d A_s^{\beta\xi}}{\d w_{\zeta,n}} 
\d_x^s \circ {\delta_\xi} \right] \Omega _{\alpha,p+1;\un,0}.
\end{align}
The sum of these expressions taken over $i+j=\ell-1$ with the coefficient $(\hbar/2) (\r_\ell)^{\mu\nu} (-1)^{i+1}$ is a part of the final formula for the operator of deformation of the bracket.



\subsection{The coefficient of $\hbar^0$}\label{sec:hbar-0}

\subsubsection{}

Observe that
\begin{align}
 - \d_x \sum_{\gamma,n} \sum_{a+b=n} &\binom{n+1}{a}\d_x^b \Omega_{\un,0;\nu,j} \d_x^a \Omega_{\mu,i;\gamma,0} \frac{\d \Omega_{\alpha,p;\beta,0}}{\d \omega_{\gamma,n}}  \\ \notag
 =& - \sum_{\gamma,n} \sum_{a+b=n} \binom{n+1}{a}\d_x^b \Omega_{\un,0;\nu,j} \d_x^a \Omega_{\mu,i;\gamma,0} \d_{\gamma,n} \d_x \Omega_{\alpha,p;\beta,0} \\ \notag
&  +  \sum_{\gamma,n} \Omega_{\un,0;\nu,j} \d_x^{n+1} \Omega_{\mu,i;\gamma,0} \frac{\d \Omega_{\alpha,p;\beta,0}}{\d \omega_{\gamma,n}} .
\end{align}
The second summand in the right hand side of this formula is equal to
\begin{align}
 \Omega_{\un,0;\nu,j}\Omega_{\mu,i;\alpha,p;\beta,0}
& = \Omega_{\un,0;\nu,j} \sum_{\gamma,n} \frac{\d \Omega_{\mu,i;\beta,0}}{\d w_{\gamma,n}} \d_x^{n+1} \Omega_{\alpha,p;\gamma,0} \\ \notag
& = \Omega_{\un,0;\nu,j} \sum_{\gamma,n} \frac{\d \Omega_{\mu,i;\beta,0}}{\d w_{\gamma,n}} \d_x^{n} 
\sum_{s\ge 1,\xi} A^{\gamma,\xi}_s \d_x^s {\delta_\xi}\Omega_{\alpha,p+1;\un,0},
\end{align}
which is a contribution to the final formula for the operator of deformation. The first summand can be rewritten in the following way.
\begin{align}\label{eq:3s}
& - \sum_{\gamma,n} \sum_{a+b=n} \binom{n+1}{a}\d_x^b \Omega_{\un,0;\nu,j} \d_x^a \Omega_{\mu,i;\gamma,0} \d_{\gamma,n} 
\sum_{s\ge 1,\xi} A^{\beta,\xi}_s \d_x^s \delta_\xi\Omega_{\alpha,p+1;\un,0} \\ \notag
& =
- \sum_{\gamma,n}\sum_{a+b=n} \binom{n+1}{a}\d_x^b \Omega_{\un,0;\nu,j} \d_x^a \Omega_{\mu,i;\gamma,0}  
\sum_{s\ge 1,\xi}\frac{\d A^{\beta,\xi}_s}{\d w_{\gamma,n}} \d_x^s \delta_\xi\Omega_{\alpha,p+1;\un,0} 
\\ \notag
&
- \sum_{s\ge 1,\xi} A^{\beta,\xi}_s \d_x^s 
\sum_{\gamma,n} \sum_{a+b=n} \binom{n+1}{a}\d_x^b \Omega_{\un,0;\nu,j} \d_x^a \Omega_{\mu,i;\gamma,0} \d_{\gamma,n}
\delta_\xi\Omega_{\alpha,p+1;\un,0} \\ \notag
&
- \sum_{s\ge 1,\xi} A^{\beta,\xi}_s \sum_{f+e=s-1} \d_x^f \left( \Omega_{\un,0;\nu,j}
\d_x^e \sum_{\gamma,n}^\infty
\d_x^{n+1} \Omega_{\mu,i;\gamma,0} \d_{\gamma,n}
\delta_\xi\Omega_{\alpha,p+1;\un,0}\right).
\end{align}
Here the first summand is a contribution to the final formula. The second summand will appear once again with the opposite sign, see the comment after equation~\eqref{eq:4s}. The third summand can be rewritten using the following computation:
\begin{align}
& \sum_{n=0}^\infty
\d_x^{n+1} \Omega_{\mu,i;\gamma,0} \d_{\gamma,n}
\delta_\xi\Omega_{\alpha,p+1;\un,0}
\\ \notag
& =
\sum_{n=0}^\infty
\d_x^{n+1} \Omega_{\mu,i;\gamma,0} \d_{\gamma,n}
\delta_\xi\Omega_{\alpha,p+1;\un,0}
- 
\delta_\xi
\sum_{n=0}^\infty
\d_x^{n+1} \Omega_{\mu,i;\gamma,0} \d_{\gamma,n}
\Omega_{\alpha,p+1;\un,0} \\ \notag
& =
- \sum_{n=0}^\infty \T_{\xi,n} \d_x\Omega_{\mu,i;\gamma,0} (-\d_x)^{n} \delta_\gamma \Omega_{\alpha,p+1;\un,0},
\end{align}
Here we used Lemma~\ref{lemma:commutator} for the last equality and the first equality comes from the observation
\begin{equation}
\delta_\xi
\sum_{n=0}^\infty
\d_x^{n+1} \Omega_{\mu,i;\gamma,0} \d_{\gamma,n}
\Omega_{\alpha,p+1;\un,0} =
\delta_\xi \Omega_{\mu,i;\alpha,p+1;\un,0}
=\delta_\xi\d_x \Omega_{\mu,i;\alpha,p+1}=0.
\end{equation}
So, the third summand of the right hand side of Equation~\eqref{eq:3s} is equal to 
\begin{equation}
\sum_{s\ge 1,\xi} A^{\beta,\xi}_s \sum_{f+e=s-1} \d_x^f \left(\Omega_{\un,0;\nu,j}
\d_x^e \sum_{\gamma,n} \T_{\xi,n}\d_x \Omega_{\mu,i;\gamma,0} (-\d_x)^{n} \delta_\gamma \Omega_{\alpha,p+1;\un,0}\right),
\end{equation}
which is again a part of final formula. 

\subsubsection{}

We have:
\begin{align} \label{eq:2sl}
& \d_x\left(\Omega_{\beta,0;\nu,j}\Omega_{\mu,i;\alpha,p}\right) = \d_x\Omega_{\beta,0;\nu,j}\Omega_{\mu,i;\alpha,p}  
\\ \notag
&  + \Omega_{\beta,0;\nu,j}\sum_{\gamma,n}\frac{\d \Omega_{\mu,i;\un,0}}{\d w_{\gamma,n}} \d_x^n \sum_{s\ge 1,\xi} A^{\gamma,\xi}_s \d_x^s \delta_\xi \Omega_{\alpha,p+1;\un,0}.
\end{align}
The second summand is a part of the final formula for the operator of deformation. The first summand is considered in section~\ref{sec:exceptional}.

\subsubsection{}

We observe that
\begin{align}\label{eq:delta1}
& \delta_\xi\left( 
\d_x^b \Omega_{\un,0;\nu,j} \d_x^a \Omega_{\mu,i;\gamma,0} \frac{\d \Omega_{\alpha,p+1;\un,0}}{\d \omega_{\gamma,n}}\right)
\\ \notag
& = \sum_{k,\ell\geq 0} (-1)^{k+l} \binom{k+l}{k} \T_{\xi,k+l} \d_x^b
\Omega_{\un,0;\nu,j} \d_x^{a+k} \Omega_{\mu,i;\gamma,0} \d_x^\ell \d_{\gamma,n} \Omega_{\alpha,p+1;\un,0}
\\ \notag
& + \sum_{k,\ell\geq 0} (-1)^{k+l} \binom{k+l}{k} \d_x^{b+k}
\Omega_{\un,0;\nu,j} \T_{\xi,k+l} \d_x^{a} \Omega_{\mu,i;\gamma,0} \d_x^\ell \d_{\gamma,n} \Omega_{\alpha,p+1;\un,0}
\\ \notag
& + \sum_{k,\ell\geq 0} (-1)^{k+l} \binom{k+l}{k} \d_x^{b+k}
\Omega_{\un,0;\nu,j} \d_x^{a+\ell} \Omega_{\mu,i;\gamma,0} \T_{\xi,k+l} \d_{\gamma,n} \Omega_{\alpha,p+1;\un,0}.
\end{align}
Meanwhile,
\begin{align}\label{eq:delta2}
& \delta_\xi \left(\Omega_{\un,0;\nu,j}\Omega_{\mu,i;\alpha,p+1}\right) =  \delta_\xi \Omega_{\un,0;\nu,j} \Omega_{\mu,i;\alpha,p+1}
\\ \notag
& + \sum_{n\geq 0}(-1)^{n+1} \T_{\xi,n+1}\Omega_{\un,0;\nu,j} \d_x^{n} \sum_{\gamma,k} \d_x^{k+1} \Omega_{\mu,i;\gamma,0} \d_{\gamma,k} \Omega_{\alpha,p+1;\un,0} 
\\ \notag
& + \sum_{n\geq 0}(-1)^n \d_x^{n}\Omega_{\un,0;\nu,j} \T_{\xi,n+1} \sum_{\gamma,k} \d_x^{k+1} \Omega_{\mu,i;\gamma,0} \d_{\gamma,k} \Omega_{\alpha,p+1;\un,0} 
\end{align}
(here we used that $\T_{\xi,n+1}\d_x = \T_{\xi,n}$). Therefore, by a direct computation of the combinatorial coefficients, we see that 
\begin{align}\label{eq:4s}
& \delta_\xi \left[ \sum_{\gamma,n} \sum_{a+b=n} \binom{n+1}{a}
\d_x^b \Omega_{\un,0;\nu,j} \d_x^a \Omega_{\mu,i;\gamma,0} \frac{\d \Omega_{\alpha,p+1;\un,0}}{\d \omega_{\gamma,n}}
-\Omega_{\un,0;\nu,j}\Omega_{\mu,i;\alpha,p+1}\right] \\ \notag
& = 
\sum_{\gamma}\sum_{0\leq u\leq v} \binom{v}{u}
\T_{\xi,v+1}\Omega_{\un,0;\nu,j} (-\d_x)^{v-u} \Omega_{\mu,i;\gamma,0} (-\d_x)^{u+1} \delta_\gamma \Omega_{\alpha,p+1;\un,0} 
 \\ \notag
& - \sum_{\gamma}\sum_{0\leq u\leq v} \binom{v+1}{u}
(-\d_x)^{v-u}\Omega_{\un,0;\nu,j} \T_{\xi,v+1} \Omega_{\mu,i;\gamma,0} (-\d_x)^{u+1} \delta_\gamma \Omega_{\alpha,p+1;\un,0} 
\\ \notag 
& + \sum_{\gamma}\sum_{u,v,n=0}^\infty (-1)^n \binom{u+v+1}{v}
(-\d_x)^{u}\Omega_{\un,0;\nu,j} (-\d_x)^v \Omega_{\mu,i;\gamma,0} \T_{\xi,u+v-n}\d_{\gamma,n} \Omega_{\alpha,p+1;\un,0} 
\\ \notag 
& + \delta_\xi \Omega_{\un,0;\nu,j} \left(\sum_{\gamma}\sum_{0\leq u\leq n} \d_x^{n-u}\Omega_{\mu,i;\gamma,0} (-\d_x)^u\d_{\gamma,n} \Omega_{\alpha,p+1;\un,0} -\Omega_{\mu,i;\alpha,p+1}\right).
\end{align}
We should apply the operator $\sum_{s=1}^\infty A^{\beta\xi}_s \d_x^s$ to this expression. The first and the second summand are parts of the final formula. The third summand will turn into the second summand in the right hand side of the equation~\eqref{eq:3s} with the opposite sign, so they will cancel each other. The fourth summand will be equal to the following:
\begin{align}
& \sum_{s=1}^{\infty} A^{\beta\xi}_s \d_x^s \left( 
\delta_\xi \Omega_{\un,0;\nu,j} \left(\sum_{\gamma}\sum_{0\leq u\leq n} \d_x^{n-u}\Omega_{\mu,i;\gamma,0} (-\d_x)^u\d_{\gamma,n} \Omega_{\alpha,p+1;\un,0} -\Omega_{\mu,i;\alpha,p+1}\right)\right) 
\\ \notag
& = \sum_{s=1}^{\infty} A^{\beta\xi}_s \d_x^s 
\delta_\xi \Omega_{\un,0;\nu,j} \left(\sum_{\gamma}\sum_{0\leq u\leq n} \d_x^{n-u}\Omega_{\mu,i;\gamma,0} (-\d_x)^u\d_{\gamma,n} \Omega_{\alpha,p+1;\un,0} -\Omega_{\mu,i;\alpha,p+1}\right) 
\\ \notag
& + \sum_{s=1}^\infty A^{\beta\xi}_s \sum_{\gamma}\sum_{e+f=s-1} \binom{s}{e} \d_x^e 
\delta_\xi \Omega_{\un,0;\nu,j} \d_x^f\left(\Omega_{\mu,i;\gamma,0} \d_x \delta_\gamma \Omega_{\alpha,p+1;\un,0}\right).
\end{align}
Here the second summand is again a contribution to the final formula, and the first summand is considered in the next section together with the first summand in the right hand side of equation~\eqref{eq:2sl}.

\subsubsection{}\label{sec:exceptional} In this section we collect all expressions that are not yet converted into the contributions to the final formula. We have:
\begin{align}
& \sum_{i+j=l-1}(-1)^{i+1} \d_x\Omega_{\beta,0;\nu,j-1} \left(\sum_{\gamma}\sum_{0\leq u\leq n} \d_x^{n-u}\Omega_{\mu,i;\gamma,0} (-\d_x)^u\d_{\gamma,n} \Omega_{\alpha,p+1;\un,0} \right. \\ \notag
& -\Omega_{\mu,i;\alpha,p+1}\left) +\sum_{i+j=l-1}(-1)^{i+1} \d_x\Omega_{\beta,0;\nu,j} \Omega_{\mu,i;\alpha,p} \right. 
\\ \notag
& = \sum_{i+j=l-1}(-1)^{i+1} \d_x\Omega_{\beta,0;\nu,j-1} X_{\mu,i},
\end{align}
where $X_{\mu,i}$ is equal to
\begin{align}
-\Omega_{\mu,i+1;\alpha,p} - \Omega_{\mu,i;\alpha,p+1} + \sum_{\gamma}\sum_{0\leq u\leq n} \d_x^{n-u}\Omega_{\mu,i;\gamma,0} (-\d_x)^u\d_{\gamma,n} \Omega_{\alpha,p+1;\un,0} 
\end{align}

We observe that 
\begin{align}\label{eq:sum}
& \Omega_{\mu,i+1;\alpha,p} + \Omega_{\mu,i;\alpha,p+1} \\ \notag
& 
= \d_x^{-1} \sum_{\begin{smallmatrix}\gamma,n \\ \xi,m \end{smallmatrix}}
\d_{\xi,m} \Omega_{\mu,i+1;\un,0} \d_x^m \sum_{s=1}^\infty A^{\xi\gamma}_s\d_x^s (-\d_x)^n \d_{\gamma,n} \Omega_{\alpha,p+1;\un,0}
\\ \notag
& + \d_x^{-1} \sum_{\begin{smallmatrix}\gamma,n \\ \xi,m \end{smallmatrix}}
\d_x^n \left(\sum_{s=1}^\infty A^{\gamma\xi}_s\d_x^s (-\d_x)^m \d_{\xi,m} \Omega_{\mu,i+1;\un,0}\right) \d_{\gamma,n} \Omega_{\alpha,p+1;\un,0}
\end{align}
Here $\d_x^{-1}$ is a formal left inverse to $\d_x$, whose main property is that for any functions $A$ and $B$ 
\begin{equation}
\d_x^{-1}\left(\d_x A \cdot B\right) = \d_x^{-1}\left( A \cdot (-\d_x) B\right) + A\cdot B.
\end{equation}
Since $\sum_{s=1}^\infty A^{\xi\gamma}_s\d_x^s$ is an operator that defines a Poisson structure on the space of local functionals, 
\begin{equation}
\sum_{s=1}^\infty A^{\xi\gamma}_s\d_x^s = - \sum_{s=1}^\infty (-\d_x)^s \circ A^{\gamma\xi}_s.
\end{equation}
Using this two observations, we can rewrite equation~\eqref{eq:sum} in the following way:
\begin{align}
& \Omega_{\mu,i+1;\alpha,p} + \Omega_{\mu,i;\alpha,p+1} \\ \notag
& 
= \sum_{\begin{smallmatrix}\gamma,n \\ \xi,m \end{smallmatrix}} \left(
\sum_{u=0}^{m-1} 
(-\d_x)^u \d_{\xi,m} \Omega_{\mu,i+1;\un,0} \d_x^{m-1-u} \sum_{s=1}^\infty A^{\xi\gamma}_s\d_x^s (-\d_x)^n \d_{\gamma,n} \Omega_{\alpha,p+1;\un,0}
\right. 
\\ \notag
& 
+ \sum_{1\leq e \leq s}^\infty 
(-\d_x)^{e-1} \left( A^{\xi\gamma}_s (-\d_x)^m \d_{\xi,m} \Omega_{\mu,i+1;\un,0} \right) \d_x^{s-e} (-\d_x)^n \d_{\gamma,n} \Omega_{\alpha,p+1;\un,0}
\\ \notag
& \left.
+ \sum_{v=0}^{n-1}
\d_x^{n-1-v} \left(\sum_{s=1}^\infty A^{\gamma\xi}_s \d_x^{s}  (-\d_x)^m \d_{\xi,m} \Omega_{\mu,i+1;\un,0} \right) (-\d_x)^v \d_{\gamma,n} \Omega_{\alpha,p+1;\un,0}\right) .
\end{align}
Therefore,
\begin{align}
& X_{\mu,i} = \sum_{\gamma}\left(
-\sum_{\xi,m} \sum_{u=0}^{m-1} 
(-\d_x)^u \d_{\xi,m} \Omega_{\mu,i+1;\un,0} \d_x^{m-1-u} \circ\sum_{s=1}^\infty A^{\xi\gamma}_s\d_x^s \right. 
\\ \notag
& \left. -\sum_{\xi}\sum_{2\leq f \leq s}^\infty 
(-\d_x)^{s-f} \left( A^{\xi\gamma}_s \delta_{\xi} \Omega_{\mu,i+1;\un,0}\right) \d_x^{f-1} \right) \delta_{\gamma} \Omega_{\alpha,p+1;\un,0},
\end{align}
and $\sum_{i+j=l-1}(-1)^{i+1} \d_x\Omega_{\beta,0;\nu,j-1} X_{\mu,i}$ is a contribution to the final formula.

This computation completes the proof of Theorem~\ref{thm:r-def-A}.

\subsection{A formula for the operator of $\s$-deformation}

In this section we prove a formula for the $\s$-deformation. 

\begin{theorem}\label{thm:s-def-A} We have: 
\begin{align}
& \sum_{s=1}^{\infty} \left(\sum_{\ell=1}^\infty\widehat{\s_\ell z^\ell}[w]. A_{s}^{\beta\xi}\right) \d_x^s 
= -\sum_{s=1}^{\infty} \left(\sum_{\gamma} (\s_1)_{\gamma,\un} \frac{\d A_{s}^{\beta\xi}}{\d w_{\gamma,0}}\right) \d_x^s.
\end{align}
\end{theorem}

\begin{proof} This is a straightforward computation. We have: 
\begin{align}
& \d_x \sum_{\ell=1}^\infty\widehat{\s_\ell z^\ell}[w]. \Omega_{\alpha,p;\beta,0}
- \sum_{\xi}\sum_{s=1}^\infty A_s^{\beta,\xi}\d_x^s \delta_\xi \sum_{\ell=1}^\infty\widehat{\s_\ell z^\ell}[w].\Omega_{\alpha,p+1;\un,0} 
\\ \notag &
= 
\sum_{1\leq\ell\leq p}(\s_\ell)^{\mu}_{\alpha}\d_x\Omega_{\mu,p-\ell;\beta,0}
-\d_x\sum_{\gamma}\frac{\partial\Omega_{\alpha,p;\beta,0}}{\partial w_{\gamma,0}} (\s_1)_{\gamma,\un}
\\ \notag &
- \sum_{\xi}\sum_{s=1}^\infty A_s^{\beta,\xi}\d_x^s \delta_\xi
\left(\sum_{1\leq\ell\leq p+1}(\s_\ell)^{\mu}_{\alpha}\Omega_{\mu,p+1-\ell;\un,0}
-\sum_{\gamma}\frac{\partial\Omega_{\alpha,p+1;\un,0}}{\partial w_{\gamma,0}} (\s_1)_{\gamma,\un}\right)
\\ \notag & 
=  \sum_{\gamma}(\s_1)_{\gamma,\un} \left(-\d_{\gamma,0}\circ \sum_{\xi}\sum_{s=1}^\infty A_s^{\beta,\xi}\d_x^s\circ \delta_\xi
+ \sum_{\xi}\sum_{s=1}^\infty A_s^{\beta,\xi}\d_x^s\circ \delta_\xi\circ \d_{\gamma,0} \right)\Omega_{\alpha,p+1;\un,0}
\\ \notag &
= -\sum_{\xi}\sum_{s=1}^{\infty} \left(\sum_{\gamma} (\s_1)_{\gamma,\un} \frac{\d A_{s}^{\beta\xi}}{\d w_{\gamma,0}}\right) \d_x^s
\delta_\xi \Omega_{\alpha,p+1;\un,0}.
\end{align}
Here we used that $\sum_{\xi}\sum_{s=1}^\infty A_s^{\beta,\xi}\d_x^s \delta_\xi\Omega_{\mu,p+1-\ell;\un,0}$ is equal to $\d_x\Omega_{\mu,p-\ell;\beta,0}$ for $1\leq\ell\leq p$ and to $0$ for $\ell=p+1$.
\end{proof}




\section{Uniqueness of the bracket} \label{sec:unique}

Consider the infinitesimal deformations of the Poisson bracket (or rather of the operator $\sum_{s=1}^\infty A_s^{\beta,\xi}\d_x^s$) obtained in the previous section. It gives us a system of vector fields on the space of all operators of that type. Consider a flow line of one of these vector fields that starts at a point corresponding to the weak quasi-Miura transformation $w_\gamma=v_\gamma+\sum_{g=1}^\infty \hbar^g \d^2F_g/\d t_{\gamma_0}\d t_{\un,0}$ of the operator $\delta^{\beta,\xi}\d_x$. In principle, though the whole flow line of operators satisfies the desired property 
\begin{align}
& \sum_{\xi}\sum_{s=1}^\infty A_s^{\beta,\xi}\d_x^s\delta_{\xi} \Omega_{\alpha,p+1;\un,0} = \d_x\Omega_{\alpha,p;\beta,0},
& \mbox{for all } \alpha, \beta, \mbox{ and } p,
\end{align} 
we still have to prove that they do coincide with the corresponding weak quasi-Miura transformations of $\delta^{\beta,\xi}\d_x$ at all points of the flow line.

First, let us apply the inverse of the weak quasi-Miura transformation.
 
\begin{lemma}
The inverse weak quasi-Miura transformation, $v_\gamma=w_\gamma-\sum_{g=1}^\infty \hbar^g \d^2F_g/\d t_{\gamma,0}\d t_{\un,0}$, 
maps an operator $\sum_{s=1}^\infty A_s^{\beta,\xi}\d_x^s$ into one that also has no constant term, that is, into an operator $\sum_{s=0}^\infty B_s^{\beta,\xi}\d_x^s$ where $B_0^{\beta,\xi}=0$.
\end{lemma}
\begin{proof}
Indeed, 
\begin{align}
\sum_{s=0}^\infty B_s^{\beta,\xi}\d_x^s & := \sum_{\begin{smallmatrix} \mu,e \\ \nu,f \end{smallmatrix}} 
\frac{\d v_\beta}{\d w_{\mu,e}} \d_x^e \circ \sum_{s=1}^\infty A_s^{\mu,\nu}\d_x^s (-\d_x)^f \circ \frac{\d v_\xi}{\d w_{\nu,f}}.
\end{align}
Therefore, $B_0^{\beta,\xi}$ is equal to 
\begin{align}
\sum_{\begin{smallmatrix} \mu,e \\ \nu \end{smallmatrix}} 
\frac{\d v_\beta}{\d w_{\mu,e}} \d_x^e\sum_{s=1}^\infty A_s^{\mu,\nu}\d_x^s \frac{\delta v_{\xi}}{\delta w_{\nu}}.
\end{align}
Since $w_\beta-v_\beta$ is equal to $\d_x G_\beta$, where $G_\beta=\sum_{g=1}^\infty \hbar^g \d F_g/\d t_{\beta,0}$ is a series in $\hbar$ whose coefficients depend only on a finite number of derivatives (both in coordinates $v$ and $w$), $\delta v_{\xi}/\delta w_{\nu} = \delta_{\xi,\nu}$. Since $\d_x^s(\delta_{\xi,\nu})$ is equal to $0$ for any $s\geq 1$, we conclude that $B_0^{\beta,\xi}=0$.
\end{proof}

Now we see that the following uniqueness in genus $0$ is sufficient.

\begin{proposition}\label{prop:unique} Any operator of the form $\sum_{s=1}^\infty B_s^{\beta,\xi}\d_x^s$ such that
\begin{align}
\d_x\Omega^{[0]}_{\alpha,p;\beta,0} = \sum_{\xi}\sum_{s=1}^\infty B_s^{\beta,\xi}\d_x^s \frac{\delta \Omega^{[0]}_{\alpha,p+1;\un,0}}{\delta v_\xi}
\end{align}
is equal to $\delta^{\beta,\xi}\d_x$.
\end{proposition}

\begin{proof} We denote by $\sum_{s=1}^\infty C_s^{\beta,\xi}\d_x^s$ the difference $\left(\sum_{s=1}^\infty B_s^{\beta,\xi}\d_x^s-\delta^{\beta,\xi}\d_x\right)$. Using the topological recursion relation in genus $0$, we observe that 
\begin{align}
& \frac{\delta \Omega^{[0]}_{\alpha,p+1;\un,0}}{\delta v_\xi} = \frac{\d \Omega^{[0]}_{\alpha,p+1;\un,0}}{\d v_\xi} =
\Omega^{[0]}_{\alpha,p;\xi,0}
\end{align}
Therefore,
\begin{align}
0 & = \sum_{\xi}\sum_{s=1}^\infty C_s^{\beta,\xi}\d_x^s \frac{\delta \Omega^{[0]}_{\alpha,p+1;\un,0}}{\delta v_\xi} 
 = \sum_{\xi}\sum_{s=1}^\infty C_s^{\beta,\xi}\d_x^{s-1} \d_x\Omega^{[0]}_{\alpha,p;\xi,0} \\ \notag
& = \sum_{\xi}\sum_{s=1}^\infty C_s^{\beta,\xi}\d_x^{s-1} \frac{\d v_{\xi,0}}{\d t_{\alpha,p}} 
 = \sum_{\xi}\sum_{s=1}^\infty C_s^{\beta,\xi} \frac{\d v_{\xi,s-1}}{\d t_{\alpha,p}}.
\end{align}
Since the change of variables $t_{\alpha,p}\leftrightarrow v_{\xi,s}$ is non-degenerate, we conclude that all coefficients $C_s^{\beta,\xi}$ are equal to zero.
\end{proof}


\section{$\hbar$-Homogeneity in the orbit} \label{sec:conclusions}

In this section, we explain the polynomiality of $\Omega_{\alpha,p;\beta,q}$ and the coefficient of the operator $A_s^{\alpha\beta}\d_x^s$ (that determines the Poisson bracket of the full hierarchy) considered as functions of $w_1,w_2,\dots$.

\begin{theorem} For any tame partition function in the Givental orbit of $Z_{KdV}^{\otimes s}$, we have the following expansion:
\begin{equation}
\Omega_{\alpha,p;\beta,q}=\sum_{g=0}^\infty \hbar^g \Omega_{\alpha,p;\beta,q}^{[g]}(w,w_1,w_2,\dots),
\end{equation}
where $\Omega_{\alpha,p;\beta,q}^{[g]}$ is a homogeneous polynomial in $w_1,\dots,w_{2g}$ of degree $2g$ (here $\deg w_i=i$).
\end{theorem}

We call below the this kind of homogeneous polynomiality, that is, homogeneous polynomiality in $\hbar$-expansion, the $\hbar$-homogeneity.

\begin{proof} We have this property at one point in the orbit --- for $s$ copies of the KdV hierarchy. See Example~\ref{KdV} above, and a full description of the KdV hierarchy in~\cite{DubZha2}. 

Let us now look at the deformation formula, given by Equations~\eqref{eq:def-Omega} and~\eqref{eq:def-Omega-low}. It is easy to see that the right hand sides of both formulas are $\hbar$-homogeneous polynomials, if all $\Omega_{\alpha,p;\beta,q}$ are. Indeed, the product of two $\hbar$-homogeneous polynomials is again an $\hbar$-homogeneous polynomial, the derivatives $\d/\d_{w\xi,k}$ decrease the degree of homogeneity by $k$, the derivatives $\d_x$ increase the degree by $1$. The last summand in the right hand side of Equation~\eqref{eq:def-Omega} is multiplied by $\hbar$, and simulteneously, its homodeneous degree is shifted by $2$.

In order to apply an element of the Givental group, that is, in order to integrate the Lie algebra action, we are to solve an ODE, whose right hand side is given by Equations~\eqref{eq:def-Omega} and~\eqref{eq:def-Omega-low}. Then a standard argument implies that if a solution of this ODE is an $\hbar$-homogeneous polynomial at one point, it remains to be an $\hbar$-homogeneous polynomial at any other point.
\end{proof}

\begin{theorem} For any tame partition function in the Givental orbit of $Z_{KdV}^{\otimes s}$, the operator that determines the Poisson bracket of the full hierarchy,
\begin{equation}
\sum_{s=1}^\infty A_s^{\alpha\beta}\d_x^s := \sum_{\begin{smallmatrix} \mu,e \\ \nu,f \end{smallmatrix}} 
\frac{\d w_\alpha}{\d v_{\mu,e}} \d_x^e \circ \d_x\circ (-\d_x)^f \circ \frac{\d w_\beta}{\d v_{\nu,f}},
\end{equation}
is $\hbar$-homogeneous in $w_1,w_2,\dots$. More precisely, 
$A_s^{\alpha\beta}=\sum_{g=0}^\infty \hbar^g A_{g,s}^{\alpha\beta}$,
where $A_{g,s}^{\alpha\beta}$ is a homogeneous polynomial in $w_1,\dots,w_{2g}$ of degree $2g-s$.
\end{theorem}

\begin{proof} The proof is the same as above. Proposition~\ref{prop:unique} imply that the deformations formulas given in Theorems~\ref{thm:r-def-A} and~\ref{thm:s-def-A} are indeed the deformation formulas for the weak quasi-Miura image of the operator $\d_x$ under the change of variables $v_\alpha\mapsto w_\alpha$. We know that for KdV, this quasi-Miura image of $\d_x$ is again $\d_x$, that is, it is indeed an $\hbar$-homogenenous polynomial in $w_1,w_2,\dots$ with the right degrees of homogeneity. Also we already know the $\hbar$-homogeneity for $\Omega_{\alpha,p;\beta,q}$. Therefore, analyzing the deformation formulas in Theorems~\ref{thm:r-def-A} and~\ref{thm:s-def-A}, we see that the right hand sides of these formulas are again $\hbar$-homogeneous polynomials of the right degree. Then the same ODE-solution argument as above implies that the bracket operator is $\hbar$-homogeneous for any point in the Givental orbit of $Z_{KdV}^{\otimes s}$.
\end{proof}

The last thing that we would like to mention is that Dubrovin and Zhang have proved that in the case of a conformal Frobenius structure, the topological tau-function of their full hierarchy always lie in the Givental orbit of $Z_{KdV}^{\otimes s}$. Therefore, in that case we always have a polynomial Poisson bracket for their hierarchy. 


\end{document}